\documentclass[aps,prd,showpacs,amssymb,float]{revtex4}
\newcommand{\be}{\begin{equation}}
\newcommand{\ee}{\end{equation}}
\newcommand{\beq}{\begin{eqnarray}}
\newcommand{\eeq}{\end{eqnarray}}
\usepackage{bm}
\usepackage{graphicx}%
\usepackage{epsf}
\usepackage{epsfig}

\usepackage{amsmath}
\usepackage{amsfonts,amssymb}
\usepackage{hhline}
\usepackage{multirow}

\newcommand{\Aslash}{\ensuremath \hspace{0.08cm} \raisebox{0.025cm}{\slash}\hspace{-0.25cm} A}
\newcommand{\dslash}{\not{\hbox{\kern-2pt $\partial$}}}
\newcommand{\Tr}{\mbox{Tr}}
\newcommand{\TrR}{\mbox{Tr}_{\cal R}}
\newcommand{\YES}{$\diamond$}
\newcommand{\NO}{$\times$}
\newcommand{\third}{third }
\newcommand{\ind}{3}

\begin{document}
\title{Discrete symmetry breaking and baryon currents in U($N$) and SU($N$) gauge theories}
\author{B.~Lucini and A.~Patella}
\affiliation{Physics Department, Swansea University, Singleton Park, Swansea SA2 8PP, UK}
\begin{abstract}
In SU($N$) gauge theories with fermions in the fundamental or in a two-index 
(either symmetric or antisymmetric) representation formulated on a manifold
with at least one compact dimension with non-trivial holonomy the discrete
symmetries $C$, $P$ and $T$ are broken at small enough size of the compact
direction(s) for certain values of $N$. We show that for those $N$ in the broken
phase a non-zero baryon current wrapping in the compact direction exists, which
provides a measurable observable for the breaking of $C$, $P$ and $T$.
We prove that in all cases where the current is absent there is no breaking
of those discrete symmetries. This includes the limit $N \to \infty$ of the
SU($N$) gauge theory with symmetric or antisymmetric fermions and U($N$) gauge
theory at any value of $N$. We then argue that the component of the baryon
current in the compact direction is the physical order parameter for $C$, $P$
and $T$ breaking due to the breaking of Lorentz invariance.
\end{abstract}
\maketitle

\section{Introduction}

An interesting problem is the realisation of charge conjugation
($C$), parity ($P$) and time reversal ($T$) symmetries for U($N$) and SU($N$) gauge theories with fermions in the fundamental, two-index symmetric or
antisymmetric and adjoint representations. These are strongly interacting theories including QCD, suspersymmetric theories, and candidates for mechanisms of electroweak symmetry breaking.
Although their phenomenology can be very different from QCD, in the literature those theories are often referred to as QCD-like theories; in this paper, we will follow this terminology.

An important result in Quantum Field Theory is the $CPT$ theorem, which states that every local theory, which is Lorentz-invariant and whose energy is bounded from below, is also invariant under the successive application of $P$, $C$ and $T$. The Vafa-Witten theorem~\cite{Vafa:1984xg} proves the conservation of
parity for QCD-like theories in infinite volume. As for charge conjugation,
there is a wealth of experimental evidence that $C$-parity is preserved in
QCD~\cite{Amsler:2008zzb}. Conservation of $C$ and $P$ together with the $CPT$ theorem implies the conservation of time reversal $T$. In the absense of a rigorous theorem, it is reasonable to assume that conservation of $C$ holds for generic U($N$) and SU($N$) gauge theories with $N_f$ fermion flavours transforming in the fundamental, two-index symmetric or antisymmetric and adjoint representations of the gauge group.  Hence, it is widely believed that in those theories the vacuum is symmetric under $C$, $P$ and $T$.

The situation is different on small volumes, where Lorentz invariance
is explicitly broken. In particular, it was noticed in~\cite{Unsal:2006pj}
that in U($N$) and SU($N$) gauge theories with fermions in the (anti)symmetric
representation on a $R^3 \times S^1$ manifold where the compact $S^1$ is closed
with periodic boundary conditions for fermions (and hence identifies a
compact spatial dimension), for a sufficiently small radius of the $S^1$, the
effective potential is minimised by values of the Wilson line wrapping the $S^1$ with non-vanishing imaginary part. By analogy with the thermal case, this Wilson line is also referred to as the Polyakov loop. In most of the cases, a non-real value for the Polyakov loop implies that not only the center symmetry (if any), but also $P$, $C$, $T$ and $CPT$ are spontaneously broken~\footnote{Note that contrary to what stated in~\cite{Unsal:2006pj}, a purely
imaginary Polyakov loop does not imply breaking of $P$, $C$ and $T$
symmetries, as firstly noted in~\cite{Myers:2009df}. This
issue will be discussed in detail in this paper.}. We will refer to all these symmetries as discrete symmetries.

Studying the properties of U($N$) and SU($N$) gauge theories under $C$, $P$ and $T$ on
compact spaces at various values of the sizes of the compact dimensions and
with various topologies is important for several reasons.
Firstly, understanding the behaviour of those theories under $C$, $P$ and
$T$ is mandatory for establishing rigorous equivalences in the large
$N$ limit between some classes of observables in two different theories like
orientifold planar
equivalence~\cite{Armoni:2003gp,Armoni:2003fb,Armoni:2004ub} (for which conservation of $C$ must be assumed~\cite{Unsal:2006pj,Armoni:2007rf}) or volume
independence of some observables~\cite{Kovtun:2007py,Unsal:2008ch}.
Then, a clear mapping of the phase structure of U($N$) and SU($N$) gauge theories with
fermions in two-index representations provides useful guidance to lattice
investigations of novel strong interactions as the underlying mechanism for
electroweak symmetry breaking~\cite{Catterall:2007yx,Shamir:2008pb,DelDebbio:2008zf,Catterall:2008qk,DeGrand:2008kx} and of orientifold planar
equivalence~\cite{Armoni:2008nq,Farchioni:2007dw}, which are still in their
infancy. 
Last but not least, an SU(3) gauge theory with fermions in the antisymmetric
representation, which is a possible formulation of QCD, also presents breaking
of $C$, $P$ and $T$. Hence, the problem of spontaneous breaking
of those discrete symmetries in small volume could potentially affect QCD. Not
surprisingly then, Ref.~\cite{Unsal:2006pj} has prompted several
analytical~\cite{Hollowood:2006cq,Lucini:2007as} and
numerical~\cite{DeGrand:2006qb,Lucini:2007as,DeGrand:2007tw} studies of the
phase structure of QCD-like theories on manifold with compact directions (see
also~\cite{HoyosBadajoz:2007ds} and more recently~\cite{Myers:2009df},
where the phases of gauge theories with fermions in the fundamental and in
two-index representations are discussed in great detail).
The numerical studies were mostly concerned with the restoration of $C$, $P$ and $T$ above a critical radius, which was investigated in detail
in~\cite{DeGrand:2006qb,DeGrand:2007tw}. In~\cite{Lucini:2007as}, it was shown
that a physical manifestation of this symmetry breaking in QCD is the existence
of a non-zero baryon current wrapping the $S^1$.

In this work, we further develop and extend the proposal
of~\cite{Lucini:2007as},
showing that the phenomenon of a persistent baryon current wrapping the
$S^1$ is present in all cases (among the considered ones)
in which there is $C$, $P$ and $T$ symmetry breaking. Viceversa, when this current is
null, there is no $C$, $P$ and $T$ symmetry breaking. This is also true in the cases
where the degenerate vacua are identified by a purely imaginary Polyakov loop:
we shall show that in those cases the apparent breaking is due to an
unphysical definition
of $C$, $P$ and $T$ symmetries; when the physical operators are used, the conservation
of those symmetries is manifest. The paper is organised as follows.
The issue of spontaneous breaking of $C$, $P$ and $T$ symmetries
for small compactification radius is briefly reviewed in
Sect.~\ref{sect:2}, where we also define our notations. In
Sect.~\ref{sect:3} we derive the expression of the baryon current in the
non-symmetric vacua, study under which conditions this current is different
from zero and discuss its physical origin.
In Sect.~\ref{sect:4}, further developing a recent suggestion
of~\cite{Myers:2009df}, the role of the Polyakov loop as an order parameter
for the breaking of $C$, $P$ and $T$ is revisited, and using
a physical definition of operators that implement $C$, $P$ and $T$, it is
shown that those symmetries are conserved when the Polyakov loop in the
degenerate vacua has a zero real (or imaginary) part.
This will enable us to prove that the existence of a persistent baryon current
is a necessary and sufficient condition for the breaking of $C$, $P$ and $T$ symmetries in all the considered cases.
Building on this result, in Sect.~\ref{sect:5} we conclude that the baryon
current is the physical order parameter for $C$, $P$ and $T$ symmetry breaking due to
the breaking of Lorentz invariance.

\section{Breaking of $C$, $P$ and $T$ symmetries at small compactification radius}
\label{sect:2}
We consider a gauge theory defined on a $R^3 \times S^1$ manifold. We will discuss both the case of periodic boundary conditions (PBC) and antiperiodic boundary conditions (ABC) for the $S^1$, whose size will be denoted by $L$. For convenience, we regularise the $R^3$ in the infrared by formulating the theory on a box of size $L_R^3 \times L$, with $L_R \gg L$ and taking the limit $L_R \to \infty$ of the results. As we will see, $L_R$ disappears from the final formulae. Since our results are independent of the boundary conditions on the directions send to infinity, we don't need to specify those boundary conditions.

The Lagrangian for a U($N$) or SU($N$) gauge theory with $N_f$ fermion flavours transforming in the representation ${\cal R}$ of the gauge group is given by
\beq 
\nonumber
\label{l0}
{\cal L} =  &-& \frac{1}{2 g^2} \mbox{Tr}\left( G_{\mu \nu} (x) G^{\mu \nu}(x) \right)
+ \sum_{l=1}^{N_f} \bar{\psi}_l (x)\left(i \dslash - \mathcal{R}[\Aslash] - m \right) \psi_l(x) \ , 
\eeq
where $G_{\mu \nu} = \partial_{\mu} A_{\nu} - \partial_{\nu} A_{\mu} + [A_{\mu},A_{\nu}]$ and the subscript $\mathcal{R}[\Aslash]$ indicates the gauge field in the representation ${\cal R}$. In this work, we will consider fermions in the fundamental, in the two-index symmetric and in the two-index antisymmetric representations.  Theories with fermions in the adjoint representation will not be investigated, since they do not present spontaneous breaking of $C$, $P$ and $T$ at small volume~\cite{Myers:2007vc,Myers:2009df,Cossu:2009sq}.

By choosing a diagonal background gauge field along the $S^1$ (identified as the \third direction)
\beq
\label{ansatz}
A_{\ind}^{bg} = \left(
\begin{array}{ccc}
\frac{v_1}{L} & & \\
& \ddots & \\
& & \frac{v_N}{L}
\end{array}
\right) \ ,
\eeq
with the constraint $\sum_{i} v_i = 0 \ \mbox{mod}(2 \pi)$ if the gauge group is SU($N$), and integrating out the non-zero modes (after gauge-fixing), the effective potential $V$ for the eigenvalues $v_i$ is defined (we are keeping implicit the integration over fermions, ghosts and auxiliary fields):
\beq
e^{i L V_3 V(v_1, \dots, v_N)} = \int_{\substack{\textrm{non-zero}\\\textrm{modes}}} \mathcal{D}A \ e^{i \left[ S(A+A^{bg}) + S_{GF}(A+A^{bg})\right] } \ \ ,
\eeq
where $S_{GF}$ is the gauge-fixing action and $V_3 = L_R^3$. If $v_i^*$ is a set of values that minimise the effective potential, then the partition function (in the Minkowskian formalism) is given by
\beq  
Z = e^{i L V_3 V(v_1^*, \dots, v_N^*)} \ .
\eeq
In the general case of degenerate vacua, the partition function does not depend on the vacuum (the minimum value of $V$ is always the same). On the contrary, the expectation value of Wilson line $W$ wrapping the $S^1$
\beq
W = P e^{i \int_0^L A_{\ind} \,d x^{\ind}}
\eeq
($P$ indicates path ordering), given by $\langle \Tr W \rangle = \sum_{i=1}^N e^{i v^*_i}$, does depend on the selected vacuum. In the formula above, it is understood that $A_{\ind}$ is computed at fixed values of the transverse coordinates; because of the translational symmetry along the $R^3$, the values of the transverse coordinates do not need to be specified.

The effective potential $V$ can be evaluated at one loop. The explicit derivation is standard and can be found for instance in~\cite{Myers:2009df}, which we follow closely in this section. $V$ can be split into 
\beq 
\label{effpot}
V = V_{Adj} + V_{\cal R} \ ,
\eeq
where $V_{Adj}$ is the contribution of the gauge (plus ghost) part of the action and $V_{\cal R}$ is the fermion contribution. The fermion contribution is given by
\beq
\label{eq:vferm}
V_{\cal R} = \frac{2 m^2 N_f}{\pi^2 L^2} \sum_{n=1}^{\infty}
\frac{(\pm 1)^n}{n^2} {\cal R}e \TrR W^n \ K_2(nLm) \ ,  
\eeq
where ${\cal R}e \TrR W^n$ indicates the real part of the trace of $W^n$ in the representation $\mathcal{R}$ and $K_2$ is the order two modified Bessel function of the second kind.
In Eq.~(\ref{eq:vferm}), the sum is weighted by $(-1)^n$ for ABC, while the plus sign refers to the PBC case.


The gauge contribution (which we do not need to write in detail) generates an attractive force for the eigenvalues.
Therefore the minimum of the effective potential $V$, Eq.~(\ref{effpot}) is obtained at
the point in which all the $v$ are equal and this common value $v^*$ minimises
the term due to the fermions. Introducing the $N$-ality of the representation
$N_{\cal R}$, which is one for the fundamental and two for the symmetric and antisymmetric two-index
representations, up to a term that does not depend on $v^*$, we get
\beq
\label{eq:vferm1}
V_{\cal R} = \frac{2 m^2 N_f}{\pi^2 L^2} \sum_{n=1}^{\infty}
\frac{(\pm 1)^n}{n^2} \mbox{dim}_{\cal R} \cos 
\left( N_{\cal R} n v^* \right) \ K_2(nLm) \ ,  
\eeq
$\mbox{dim}_{\cal R}$ being the dimension of the representation. The minimisation
of the effective potential must keep into account the constraint on the phases in the case of an SU($N$) gauge group. The minima in the cases of interest are listed in Table~\ref{table1}. The expectation value of the Wilson line is given by
\beq
\langle \Tr W \rangle = N e^{i v^*}
\ .
\eeq
For ABC the system is in the usual thermal phase: the center symmetry (if any) is spontaneously broken, while $P$, $C$ and $T$ are unbroken. However for PBC, also the symmetries $P$, $ C$ and $T$, defined in the standard way:
\beq
\langle \Tr W \rangle \to \langle \Tr W^{\dag} \rangle \ ,
\eeq
are broken for fundamental fermions and SU($N$) gauge group with $N$ odd, or with fermions in the (anti)symmetric two-index representations and both SU($N$) and U($N$) gauge groups with any $N$. We will discuss this definition of $P$, $C$ and $T$ in Sect.~\ref{sect:4}.

\begin{table}
\begin{center}
\begin{tabular}{|c|c|c|c|c|c|}
\hline
BC & $\mathcal{R}$ & Group & Vacua & $\langle \Tr W \rangle \to \langle \Tr W^\dagger \rangle $ & baryon current\\
\hline
\multirow{4}{2cm}{\centering ABC} & F                                  & SU($N$) / U($N$)          &
$ v^* = 0$ & \YES & \NO \\
                                  \cline{2-6}
                                  & \multirow{3}{2cm}{\centering S/AS} & SU($N$) with $N$ even     &
\multirow{2}{3cm}{\centering $ v^* = 0,\ \pi$} & \multirow{2}{1cm}{\centering \YES} & \multirow{2}{1cm}{\centering \NO} \\
                                  &                                    & U($N$)                    & 
                                  & & \\
                                                                       \cline{3-6}
                                  &                                    & SU($N$) with $N$ odd      &
$v^* = 0$ & \YES & \NO \\
\hline
\multirow{8}{2cm}{\centering PBC} & \multirow{3}{2cm}{\centering F}    & SU($N$) with $N$ even     &
\multirow{2}{3cm}{\centering $ v^* = \pi$} & \multirow{2}{1cm}{\centering \YES} & \multirow{2}{1cm}{\centering \NO} \\
                                  &                                    & U($N$)                    & 
                                  & & \\
                                                                       \cline{3-6}
                                  &                                    & SU($N$) with $N$ odd      &
$v^* = \frac{N \pm 1}{N}\pi$      & \NO & \YES \\
                                  \cline{2-6}
                                  & \multirow{5}{2cm}{\centering S/AS} & SU($N$) with $N=0 \mod 4$ &
\multirow{2}{3cm}{\centering $ v^* = \pm \frac{\pi}{2}$} & \multirow{2}{1cm}{\centering \NO} & \multirow{2}{1cm}{\centering \NO} \\
                                  &                                    & U($N$)                    & 
                                  & & \\
                                                                       \cline{3-6}
                                  &                                    & SU($N$) with $N=1 \mod 4$ &
$ v^* = \pm \frac{N-1}{2N}\pi $ & \NO & \YES \\
                                                                       \cline{3-6}
                                  &                                    & SU($N$) with $N=2 \mod 4$ &
$ v^* = \pm \frac{N\pm 2}{2N}\pi $ & \NO & \YES \\
                                                                       \cline{3-6}
                                  &                                    & SU($N$) with $N=3 \mod 4$ &
$ v^* = \pm \frac{N+1}{2N}\pi $ & \NO & \YES \\
\hline
\end{tabular}
\end{center}

\caption{Summarising the minima (\textit{vacua}) of the effective potential for all the gauge groups (\textit{gauge}), fermion representations ($\mathcal{R}$) and boundary conditions (\textit{BC}). The fifth column displays the properties of the vacua under the action of naively defined $P$, $C$ and $T$ (see Sect.~\ref{sect:4}); the symbols \YES{} and \NO{} indicate respectively that these symmetries are unbroken or spontaneously broken. The sixth column indicates whether a baryon current is present (\YES) or not (\NO). }
\label{table1}

\end{table}

\section{Derivation of the current}
\label{sect:3}
The Noether 4-current associated to the baryon number is
\begin{equation}
j^\mu(x) = \bar{\psi}\gamma^\mu \psi(x) \ .
\end{equation}
We showed in~\cite{Lucini:2007as} (see also~\cite{Lucini:2007im,Lucini:2007hp}) that the component $j^\ind(x)$ along the compact dimension is the local order parameter for the breaking of $P$, $C$ and $T$ in the case of SU($N$) with fermions in the fundamental representation. The existence of a non-null baryon current wrapping the $S^1$ is a physical consequence of the non-invariance
of the degenerate ground states of the system under $C$, $P$ and $T$ symmetries. Here we want to generalize this result to SU($N$) or U($N$) with fermions in the two-index symmetric and antisymmetric representations. We stress that a non-zero value for the baryon current indicates the simultaneous breaking of $P$ (and in particular the reflection symmetry with respect to a spatial plan orthogonal to the compact dimension), $C$ and $T$. In other words it is enough that one of these symmetries is conserved for the baryon current to be zero.

The vacuum expectation value of the baryon current can be defined by adding a source term in the action:
\begin{equation}
\langle j^{\ind} \rangle = - \frac{i}{Z V_3 N_f} \left. \frac{d}{d\alpha} \right|_{\alpha=0} \int {\cal D} A {\cal D} \psi {\cal D} \bar{\psi} \ \exp \left\{ iS_{YM}[A] + i \int \sum_{l=1}^{N_f} \bar{\psi}_l (x) \gamma^\nu \left( i \partial_\nu - {\cal R}[A_\nu] - m + \frac{\alpha}{L} \delta^\ind_\nu \right) \psi_l(x) \ d^3x \right\} \ .
\end{equation}
For a generic representation $\cal R$ with $N$-ality equal to $N_{\cal R} \neq 0$ the source term can be absorbed by a shift of the gauge field along the compact dimension (thanks to the invariance of the YM action under a constant shift of the gauge field):
\begin{equation} \label{eq:shift}
A_{\ind} \to A_{\ind} + \frac{\alpha}{N_{\cal R} L} \ .
\end{equation}
This argument leads to the exact formula which relates the baryon current to the effective potential of the eigenvalues of the Wilson line:
\begin{equation}
\label{formalj}
\langle j^{\ind} \rangle = \frac{L}{N_f} \left. \frac{d}{d\alpha} \right|_{\alpha=0} V\left(v_1^*+\frac{\alpha}{N_{\cal R}},\dots,v_N^*+\frac{\alpha}{N_{\cal R}}\right)
\ .
\end{equation}
The one-loop value of the baryon current can be computed by plugging the one-loop effective potential~(\ref{eq:vferm}) into~(\ref{formalj}), with the shift~(\ref{eq:shift}) implemented as $W \to \exp \left( i \frac{\alpha}{N_{\cal R}} \right) W$:
\begin{eqnarray}
\langle j^{\ind} \rangle &=& \frac{2 m^2}{\pi^2 L} \left. \frac{d}{d\alpha} \right|_{\alpha=0} \sum_{n=1}^{\infty}
\frac{(\pm 1)^n}{n^2} {\cal R}e \left( e^{i\alpha n}\TrR W^n \right) \ K_2(nLm) = \\
&=& - \frac{2 m^2}{\pi^2 L} \sum_{n=1}^{\infty}
\frac{(\pm 1)^n}{n} {\cal I}m \TrR W^n \ K_2(nLm) = \\
&=& - \frac{2 m^2}{\pi^2 L} \sum_{n=1}^{\infty}
\frac{(\pm 1)^n}{n} \sin \left( N_{\cal R} n v^* \right) \ K_2(nLm)\ .
\end{eqnarray}
The previous relationship shows that $\langle j^{\ind} \rangle \ne 0$
if  $N_{\cal R} v^* \ne k \pi/2$, with $k$ integer. Hence at sufficiently
small $L$ such that the one-loop calculation is reliable, a baryon current is generated only for gauge group SU($N$) and fermions either in the fundamental representation with PBC for odd values of $N$ or in the (anti)symmetric two-index representation and PBC for values of $N$ that are not multiple of 4 (see summary in Table~\ref{table1}). A lattice calculation in SU(3) gauge theory with fermions in the fundamental representation showed the persistence of this current beyond the perturbative regime~\cite{Lucini:2007as}.

A quick inspection of the vacua shows that at large-$N$ the SU($N$) and U($N$) gauge groups coincide, and therefore the baryon current vanishes.

A simple argument can elucidate the physical origin of the baryon current. We start from the thermal effective potential $V^{ABC}(v;\mu)$ in the presence of a chemical potential $\mu$ ($\ind$ is now the thermal direction). Even if the argument proposed here is completely non-perturbative, we quote the one-loop potential from~\cite{Myers:2009df}:
\beq
\label{eq:vfermmu}
V_{\cal R}^{ABC}(v;\mu) = \frac{m^2 N_f}{\pi^2 L^2} \sum_{n=1}^{\infty}
\frac{(-1)^n}{n^2} \left(e^{n \mu L} \TrR W^{\dag n}+ e^{- n \mu L} \TrR W^{n} \right) \ K_2(nLm) \ .
\eeq
The effective potential for PBC is obtained at imaginary chemical potential $\mu = i \pi/L$. This can be easily understood by the one-loop formula~(\ref{eq:vfermmu}), noticing that in this case the imaginary chemical potential gives an extra $(-1)^n$ factor in the sum. From a non-perturbative point of view, if $\psi$ is the fermion field with ABC, then the fermion field $\psi'$ with PBC is given by the following change of variables in the functional integral:
\begin{equation}
\psi'(x) = e^{\frac{i x^\ind \pi}{L}} \psi(x) \ .
\end{equation}
The extra phase generates in the Euclidean Lagrangian the term $i \frac{\pi}{L} \bar{\psi} \gamma^\ind \psi$, which is exactly an imaginary chemical potential.

Moreover, also the source $\alpha$ couples to the operator $i \bar{\psi} \gamma^\ind \psi$ (which is Hermitean in the Euclidean). The effective potential for PBC fermions in presence of this esternal source is:
\beq
V^{PBC}\left(v+\frac{\alpha}{N_\mathcal{R}};\mu \right) = V^{ABC}\left( v;\mu+i \frac{\pi-\alpha}{L} \right) \ .
\eeq
Here we are using that the effective potential is analitical in the chemical potential, i.e. it depends on the complex chemical potential $\mu$ but not on its complex conjugate $\bar{\mu}$. This fact is clear for the one-loop formula~\eqref{eq:vfermmu}, and it is discussed in~\cite{Roberge:1986mm} for the non-perturbative case.

The baryon current for PBC fermions is therefore computed:
\beq
\langle j^\ind \rangle_{PBC} = \frac{L}{N_f} \left. \frac{d}{d\alpha} \right|_{\alpha=0} V^{PBC}\left(v^*+\frac{\alpha}{N_{\cal R}} ; 0 \right) = - \frac{i}{N_f} \left. \frac{d}{d\mu}\right|_{\mu=i\pi/L} V^{ABC}\left( v^* ; \mu \right) \ .
\eeq
The baryon current along a spatial direction for PBC fermions is mapped into the baryon density for a thermal system (ABC) with an imaginary chemical potential.
The properties of the latter system
have been firstly studied in~\cite{Roberge:1986mm} and are one of the key
aspects of contemporary lattice investigations of systems at finite baryon
density~\cite{deForcrand:2002ci,Delia:2002gd}. In particular, it is well-known
that an imaginary chemical potential $i \pi/L $ induces a finite baryon
density in the SU(3) theory.
For U($N$) gauge theories the baryon number is zero, since it is the Noether charge of the local U(1) gauge symmetry, and the imaginary chemical potential can be gauged away.

\section{Physical definition of $C$, $P$ and $T$ symmetries}
\label{sect:4}

In Sect.~\ref{sect:2}, we have discussed the argument of~\cite{Unsal:2006pj} about the spontaneous breaking of $C$, $P$ and $T$ symmetries in QCD-like theories on manifolds with non-trivial holonomy and the proposal of using the trace of the Wilson loop wrapping the $S^1$ as an order parameter for this breaking. In Sect.~\ref{sect:3} we have shown that in most cases in which the Polyakov loop has a non-zero imaginary part a baryon current wrapping the short circle is generated. However for SU($N$) theories with $N$ multiple of 4 and U($N$) theories for arbitrary value of $N$ with the fermions in the (anti)symmetric two-index representation, the current is zero but the imaginary part of $\langle \Tr W \rangle$ is different from zero (see summary in Table~\ref{table1}). One is led to conclude that the baryon current is not a good order parameter for $C$, $P$ and $T$ breaking.

However the definition of $C$, $P$ and $T$ on the Hilbert space is not uniquely determined. In the easy case of QED, all these transformations are defined up to a phase. For generic phases for the gauge field, these transformations are not symmetries (i.e. they do not commute with the Hamiltonian). However since the transition probabilities are independent of these phases, it is enough to find one set of values for which the operators on the Hilbert space are symmetries, in order to conclude that the physics is invariant under $C$, $P$ and $T$.

For the cases in Table~\ref{table1} in which the baryon current is zero but the Wilson line has a non-zero imaginary part, we propose the following interpretation: the breaking of $C$, $P$ and $T$ is only apparent and is due to a bad definition (we will refer to it as the \textit{naive} definition) of the corresponding operators on the Hilbert space. In all these cases, an alternative definition (we will refer to it as the \textit{physical} definition) is possible. We will show that all the local observables transform in the same way, independently of the definition. In particular the physical operators still commute with the Hamiltonian, and define unbroken symmetries (unlike the naive operators). As a consequence, the baryon current wrapping the $S^1$ is shown to be the order parameter for $C$, $P$ and $T$ symmetry breaking, in all the considered cases.

As suggested in~\cite{Myers:2009df}, in either SU($N$) theories with $N$ multiple of 4 or U($N$) theories for arbitrary value of $N$, with fermions in the (anti)symmetric two-index representation, the free energy is independent of the (periodic or antiperiodic) boundary conditions for the fermions. This statement can be extended to all gauge invariant observables, which are local in the sense that they depend on the elementary fields in a region not wrapping around the compact dimension. In fact, the two vacua corresponding respectively to PBC and APB for the fermions are connected by a large gauge transformation $\Omega(x^\ind)$, which is periodic in the compact dimension up to a phase:
\beq \label{eq:large_gt}
\Omega(x^\ind+L) = \exp \left( i \frac{\pi}{2} \right) \Omega(x^\ind) \ .
\eeq
Notice that the $e^{i \pi/2}$ is always an element of the gauge group if this group is U($N$); in the case of SU($N$), the phase is an element of the group only if $N$ is a multiple of 4. In those cases, the action is invariant under the transformation~\ref{eq:large_gt}; however, PBC fermions $\psi$ in the (anti)symmetric two-index representation are mapped into ABC fermions $\psi'$:
\beq
\psi'(x^\ind+L) = \mathcal{R}[\Omega(x^\ind+L)] \psi(x^\ind+L) = \exp \left( i \pi \right) \mathcal{R}[\Omega(x^\ind)] \psi(x^\ind) = - \psi'(x^\ind) \ .
\eeq

We call $\mathcal{G}$ the unitary operator, implementing the large gauge transformation~(\ref{eq:large_gt}) on the Hilbert space:
\begin{equation}
\left| 0 \right>_{PBC} = \mathcal{G} \left| 0 \right>_{ABC} \ .
\end{equation}
If $\mathcal{O}$ is a local gauge-invariant observable (like for instance the baryon current), it is clearly not sensitive to the periodicity violation in the definition of $\mathcal{G}$, therefore $\mathcal{G} \mathcal{O} \mathcal{G}^\dagger = \mathcal{O}$ and, as anticipated:
\begin{equation}
\left< 0 \right| \mathcal{O} \left| 0 \right>_{ABC} = \left< 0 \right| \mathcal{G} \mathcal{O} \mathcal{G}^\dagger \left| 0 \right>_{PBC} = \left< 0 \right| \mathcal{O} \left| 0 \right>_{PBC} \ .
\end{equation}
The existence of the unitary operator $\mathcal{G}$ in particular implies that the thermodynamics and the spectra corresponding to different boundary conditions are exactly the same. This clearly suggests that the spontaneous breaking of $P$, $C$ and $T$ is only apparent and is due to a wrong implementation of the physical symmetries as (anti)unitary operators on the Hilbert space. The naive definitions
\begin{eqnarray}
P_0 &:& A_\mu(t,\mathbf{x}) \to g^{\mu\mu} A_\mu(t,-\mathbf{x}) \\
&& \psi(t,\mathbf{x}) \to \gamma_0 \psi(t,-\mathbf{x}) \ , \\
C_0 &:& A_\mu(t,\mathbf{x}) \to - A_\mu(t,\mathbf{x})^T \\
&& \psi(t,\mathbf{x}) \to i\gamma_0\gamma_2 \bar{\psi}(t,\mathbf{x})^T \ , \\
T_0 &:& A_\mu(t,\mathbf{x}) \to g_{\mu\mu} A_\mu(-t,\mathbf{x}) \\
&& \psi(t,\mathbf{x}) \to -i \gamma_5 \gamma_0 \gamma_2 \psi(-t,\mathbf{x}) \\ \nonumber
\end{eqnarray}
lead to unbroken symmetries in the case of ABC. For PBC the naive operators should be replaced by:
\begin{eqnarray}
&& P = \mathcal{G} P_0 \mathcal{G}^\dagger \ , \\
&& C = \mathcal{G} C_0 \mathcal{G}^\dagger \ , \\
&& T = \mathcal{G} T_0 \mathcal{G}^\dagger \ . \\ \nonumber
\end{eqnarray}
On local gauge-invariant observables, these operators and the naive ones acts in the same way; for instance $C \mathcal{O} C^\dagger = C_0 \mathcal{O} C^\dagger_0$. However the physical operators define unbroken symmetries for fermions with PBC:
\begin{equation}
C \left| 0 \right>_{PBC} = \mathcal{G} C_0 \left| 0 \right>_{ABC} = \left| 0 \right>_{PBC} \ ,
\end{equation}
and similarly for $P$ and $T$. One can actually think that the physical and naive operators represent different and independent symmetries, some of which are unbroken and some others are not. This is not true, since all the naive operators can be written as a product of the corresponding physical operator and a gauge transformation periodic up to a phase $\exp \left( i \pi \right)$, which is an element of the center $Z_2$. In the case of the charge conjugation for instance $C_0 = C \mathcal{G}^2$. This argument reveals that the broken symmetry is always and only the center symmetry.

\section{Conclusions}
\label{sect:5}
In this paper, we have shown that in SU($N$) gauge theories with fermions in the fundamental, two-index antisymmetric and two-index symmetric representations, whenever the breaking of Lorentz invariance due to the existence of a small $S^1$ induces the simultaneous breaking of $C$, $P$ and $T$ symmetries, a non-zero baryon current is induced that wrap the compact direction. Contrary to the imaginary part of the Wilson loop wrapping the $S^1$, in the case in which there are fermions in a two-index representation, the baryon current is not sensitive to the breaking of the $\mathbb{Z}_2$ symmetry. Hence, this baryon current is the true order parameter for the symmetry breaking. It would be interesting to study what happens for other gauge groups. For instance, if the gauge group is SO($N$), the theory is real and there should not be any breaking of $C$, $P$ and $T$. Accordingly, we would expect a null baryon current in that case. Theories with gauge group SO($N$) and Sp($N$) and gauge theories based 
 on exceptional groups are currently under investigation. The results will be reported in a future publication.

\section*{Acknowledgments}
We thank A. Armoni and J. Myers for helpful discussion on the physical meaning
of $C$-parity breaking. This work has been partially supported by STFC under
contracts PP/E007228/1 and ST/G000506/1. B.L. is supported by the Royal Society. 

\bibliographystyle{apsrev}
\bibliography{bcurr}
\end{document}